\documentclass[10pt]{iopart}

\usepackage{epsfig}  
\begin{document}

\title[A random walk description of the heterogeneous glassy 
dynamics of attracting colloids]{A random walk description of the 
heterogeneous glassy dynamics of attracting colloids}

\author{Pinaki Chaudhuri$^1$, Yongxiang Gao$^2$, Ludovic Berthier$^1$, 
Maria Kilfoil$^2$, and Walter Kob$^1$}

\address{$^1$ Laboratoire des Collo{\"\i}des, Verres
et Nanomat{\'e}riaux, UMR 5587, Universit{\'e} Montpellier II and CNRS,
34095 Montpellier, France}

\address{$^2$ Department of Physics, 
McGill University, Montr\'eal, Canada H3A 2T8}

\ead{berthier@lcvn.univ-montp2.fr}

\begin{abstract}
We study the heterogeneous dynamics of attractive colloidal particles 
close to the gel transition using confocal microscopy experiments
combined with a 
theoretical statistical analysis.
We focus on 
single particle dynamics and show that the
self part of the van Hove distribution function is not the 
Gaussian expected for a Fickian process, 
but that it reflects instead the existence, 
at any given time, of colloids with widely 
different mobilities. Our confocal microscopy 
measurements can be described well by a simple analytical model
based on a conventional continuous time random walk
picture, as already found in several other glassy materials. In particular, 
the theory successfully accounts for the presence of broad tails in the 
van Hove  distributions that exhibit exponential, 
rather than Gaussian, decay at large distance.
\end{abstract}

%\maketitle

\section{Dynamic heterogeneity in colloidal gels}

There are many systems in nature whose dynamics become slow
in some part of their phase diagram, because they undergo
a transition from a fluid to a disordered solid phase ---
like in a sol-gel transition, a glass transition, or a jamming 
transition. These systems are generically called ``glassy materials'',
%We think here of 
examples of which are simple or polymeric liquids,
colloidal particles with soft-core or hard-core interactions, grains, etc. 
As physicists, we would like to have a microscopic understanding of the slow
dynamics of these materials and would like to answer, in particular, 
an apparently very simple question: How do particles 
move in a glassy material close to the fluid-solid transition? 
To answer this question 
directly, one needs to resolve the dynamics of individual 
particles. In experiments, this is a particularly hard task 
for molecular liquids, although some techniques are now 
available~\cite{single,afm} 
but becomes much easier in the colloidal and granular worlds, where
direct visualization is 
possible~\cite{kegel,weeks,marty,virgile,laura,durian,gao,dibble}. 
Of course, resolving 
single particle dynamics is trivial in computer simulations 
where, for each particle in the system, the equations of motion
are directly integrated.

Hence, single particle dynamics have 
now been well documented, both numerically and experimentally, 
in a wide variety of materials. 
A most striking feature emerging from these studies 
is the existence of dynamic heterogeneity~\cite{ediger}. 
In terms of single particle trajectories, dynamic
heterogeneity implies the
existence of relatively broad distributions of mobilities 
inside the system. It is 
therefore an important task to suggest a framework 
to describe and interpret those data, and hopefully understand
the physical content carried by single particle displacements.

In this work, we study an assembly of 
moderately attractive colloidal particles (attraction depth 
$U \approx 3 k_BT$, 
where $k_BT$ is the thermal energy)
that undergo dynamic arrest at an ``intermediate'' volume fraction, 
$\phi_c \sim 0.44$~\cite{gao}.
The system is in fact  intermediate between fractal gels
made of very strongly attractive particles ($U \gg k_B T$)
at very low volume fraction, and hard sphere glasses
obtained with no attraction ($U \approx 0$) 
at a much higher volume fraction, $\phi \approx 0.6$. 
Although experiments clearly detect the presence
of an amorphous phase with arrested dynamics, the nature
of the transition towards this ``dense gel'' (or low density glass!)
remains unclear~\cite{zac}. The transition seems too far from the so-called
``attractive glass'' obtained at higher volume fraction 
in colloids with very short-range attraction (sticky particles), so that
other phenomena are usually invoked. A popular hypothesis is 
that gelation is in fact a non-equilibrium phenomenon 
due to a kinetically arrested phase separation~\cite{zac,dave}.
Dynamic heterogeneity in such systems has been analyzed before 
in just a few systems,  both experimentally~\cite{gao,dibble} and 
numerically~\cite{puertas,pablo}.

In this paper, we analyze single particle dynamics on the approach
to the glassy phase and show that the 
self-part of the van Hove distribution function is not the 
Gaussian expected for a Fickian process, 
but that it reflects instead the existence, 
at any given time, of colloids with widely 
different mobilities: Our system is dynamically heterogeneous. 
We then show that the simple analytical model proposed 
in Ref.~\cite{pinaki} to describe data in a variety of systems
close to glass and jamming transitions also describes our experimental
data in a satisfactory manner. 

This paper is organized as follows. In Sec.~\ref{exp}
we describe the system, experimental techniques, 
and the results obtained for the van Hove function. In 
Sec.~\ref{modelsec} we describe the model used to fit the experimental
data and discuss the results. We conclude the paper in 
Sec.~\ref{conclusion}.

\section{Measuring single particle dynamics using confocal microscopy}
\label{exp}

\subsection{Experimental system and techniques}

The experimental system under study is a suspension of colloidal
particles interacting through a hard-core repulsion and a softer
attractive interaction, induced by depletion
by adding polymers. Details of the system
have been presented in Ref.~\cite{gao}.
The dynamics of this system is observed using confocal fluorescence microscopy.
The strength of the inter-particle attractive interaction, $U$, is 
determined by the concentration 
of polymers in the suspension. We present data for a sample at a 
moderate
interaction strength of $U \approx 2.86 k_B T$.
We work at constant temperature $T$, so that our 
control parameter is the volume fraction of the particles, $\phi$.
We find that the system becomes a gel when $\phi$ is increased, 
with a transition close to $\phi_c \approx 0.442$~\cite{gao}.
Measurement of different relevant statistical quantities are carried 
out at different $\phi < \phi_c$. 

Our procedure to vary slowly the volume fraction uses 
particle sedimentation.
The relative buoyancy of the colloids is  
$\Delta \rho = 0.011$~g/cm$^3$, corresponding to a
gravitational height of $h = k_B T /(\frac{4}{3}\pi a^3 \Delta \rho g) 
\approx 40$ particle radii $a$, where $g$ is the acceleration due 
to gravity. Therefore, the gravitational
field is small enough that it induces a very slow 
densification of the system. The densification is slow enough
that microscopic dynamics of the colloids remains 
controlled by the interplay between attraction
and steric hindrance, rather than by sedimentation itself. 
Moreover, the large asymmetry between polymer coil diffusion 
time, $\approx 0.3$~s, 
and particle sedimentation time over one particle, $\approx 260$~s, 
ensures that polymers are uniformly distributed, maintaining 
the interaction strength $U$ constant in the course 
of the experiment.
 
\begin{figure}
\begin{center}
\psfig{file=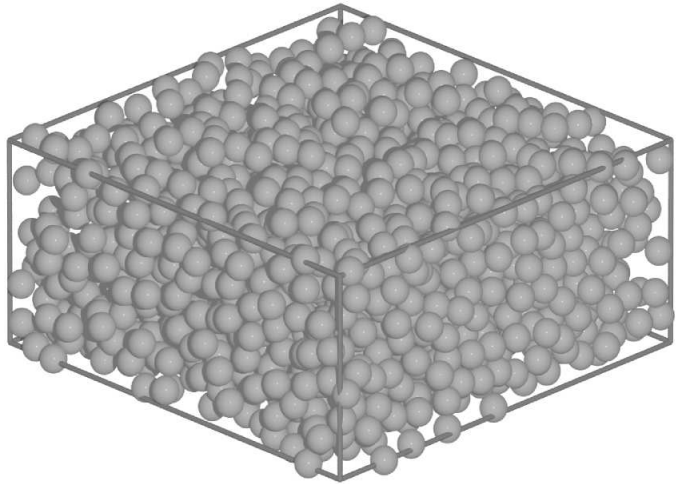,width=9.cm}
\vspace*{0.2cm}
\caption{Three-dimensional confocal microscopy rendered 
image of a typical particle
configuration at volume fraction $\phi=0.429$.}
\label{snap429}
\end{center}
\end{figure}

The colloidal particles are polymethyl-methacrylate (PMMA) spheres
of diameter $1.33~\mu$m, sterically stabilized by
chemically grafted poly-12-hydroxystearic acid, 
dyed with the electrically neutral fluorophore 4-chloro-7-nitrobenzo-2 
oxa-1,3-diazole (NBD), and suspended in a solvent mixture of 
decahydronaphthalene (decalin), tetrahydronaphthalene (tetralin), 
and cyclohexyl bromide (CXB) 
that allows for independent control of the refractive index
and buoyancy matching with the particles.
Polystyrene polymers
(molecular weight $11.4{\times}10^6$~g/mol) are added at $1.177$ mg/ml
to induce a depletion attraction at a range estimated by 
$\Delta =  2 R_g = 0.28a$, where $R_g$ is the polymer radius of
gyration. 

Using confocal microscopy,
we collect stacks of images at fixed time intervals ranging 
from $12$ to $1500$~s at different $\phi$ to access short and
long time dynamics during the approach to gelation. 
From the stacks of images we extract the particle positions of
$10^3$ particles in three dimensions and track their 
positions at better than 10~nm resolution over time.
A three-dimensional rendering of a typical
particle configuration from a stack of images 
at $\phi=0.429$ is illustrated in Fig.~\ref{snap429}.

\subsection{Non-Gaussian distributions of single particle displacements}

In an earlier work~\cite{gao}, we analyzed some structural 
and dynamical properties of the system for different volume fractions. 
In particular, we analyzed in some detail 
the distinct part of the van Hove function, finding
dynamic signatures typical of gel systems. We only presented
briefly some preliminary results concerning the self-part
of the van Hove function. It is the latter that we investigate
in more detail here. It is defined by 
\begin{equation}
G_s(x,t) = \frac{1}{N} \sum_{i=1}^N \delta (x - [x_i(t)-x_i(0)] ),
\label{gsdef}
\end{equation}
where $x_i(t)$ denotes the position of particle $i$ at time $t$ 
along one of the horizontal directions.
The function $G_s(x,t)$ measures the
probability that a given particle has undergone a displacement $x$ 
in a time interval of duration $t$.

Once the distribution (\ref{gsdef}) is known, several 
quantities can be determined. Perhaps the simplest one 
is the mean squared displacement, $\langle x^2 \rangle$, 
where the average is taken over the distribution $G_s(x,t)$, 
which contains quantitative information about the average 
mobility of the colloidal particles. In particular, its 
long-time limit yields values for the self-diffusion 
constant $D_s$ of the particles through $\langle x^2 \rangle \sim 
2 D_s t$ for large $t$.  Such a measurement, however, tells nothing about 
the possible presence of dynamic heterogeneity in the system.

\begin{figure}
\begin{center}
\psfig{file=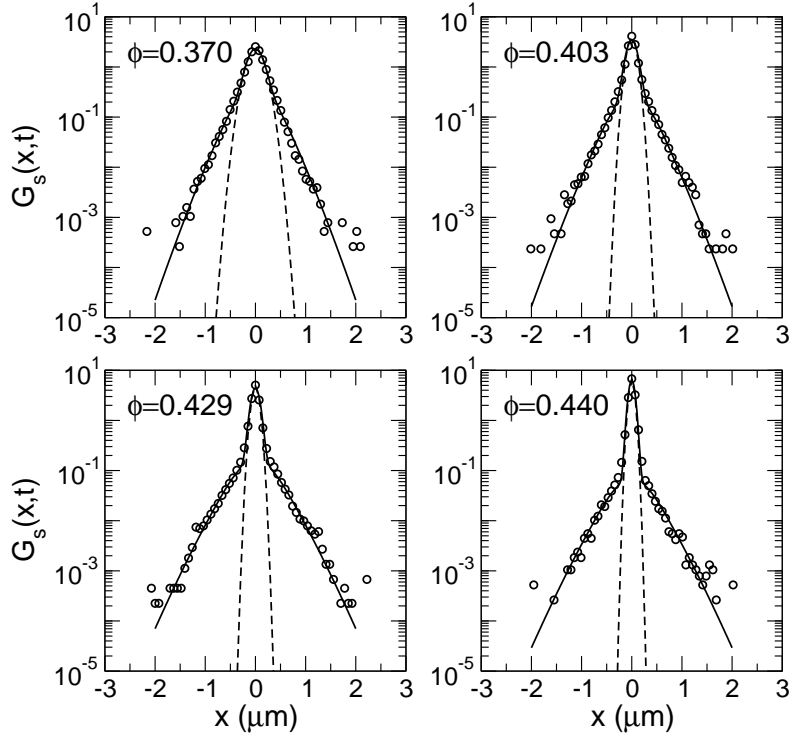,width=10.cm,angle=-90}
\end{center}
\caption{\label{fit} The self-part of the van Hove function, 
Eq.~(\ref{gsdef}),
measured using confocal microscopy upon approaching the colloidal 
gel transition by increasing the volume fraction $\phi$. For each 
$\phi$ we show the distribution for a time corresponding to average
particle displacements being close to the particle radius.
%approximately to the peak of the non-Gaussian parameter. 
Dashed lines represent Gaussian fits to the center of the distribution.
The full lines through the data are fits obtained from the 
model and parameters 
described in Sec.~\ref{modelsec}, showing very good agreement with the data.}
\end{figure}

In our earlier study~\cite{gao}, we had measured $\langle x^2 \rangle$;
although we observed a slowing down of the dynamics, the heterogeneous
nature remains hidden and could only be seen by measuring $G_s(x,t)$.
In Fig.~\ref{fit}, we have 
plotted $G_s(x,t \sim t^\star)$ for four different volume 
fractions ranging from $0.37$ to $0.44$, where $t^\star$
corresponds to the time when $\sqrt{\langle{x^2}\rangle} \approx 0.2a$
($a$ being the particle radius) - 
which is a meaningful measure of the timescale
for structural relaxation~\cite{gao}. We can clearly see that $G_s(x,t)$,
at this timescale, is very non-Gaussian and therefore, the
particle trajectories do not correspond to Fickian dynamics.

It can be easily seen that although, in all four cases, 
most of the statistical weight of the functions is carried 
by particles which have barely moved, 
$x < 0.5 \mu$m, there is a pronounced tail extending to 
distances that are much larger than what is expected 
for the Gaussian prediction shown as dashed lines \cite{gao}.
The small $x$ behavior, however, is not far from a Gaussian distribution, 
corresponding to quasi-harmonic vibrations in the cage formed by neighbouring 
particles, but at large distances the decay is well described 
by an exponential, rather than a Gaussian, decay. 
Thus, the single particle motion for our 
experimental system at timescales corresponding to $t^\star$ 
is strongly non-diffusive. Such a
non-Gaussian behaviour has been observed in other 
glass-forming systems, both in simulations~\cite{kob,stariolo,odagaki}
and experiments~\cite{kegel,weeks,marty,virgile,laura}.

Such non-Gaussianity is related
to the presence of heterogeneity in the dynamics of the 
particles in the system. Most of the particles simply undergo 
vibrational motion around their initial position ---
this corresponds to the central 
Gaussian part in $G_s(x,t)$. Additionally, 
a small fraction of the particles gets
the opportunity to explore larger distances during the 
observational time and contributing to the 
non-trivial tail of $G_s(x,t)$~\cite{kob}. Moreover, Fig.~\ref{fit} 
shows that, with increasing volume fraction, the 
width of the central Gaussian of $G_s(x,t)$ at time $t^\star$ 
decreases while 
the tail gets more pronounced. This implies that even though the 
volume available for the quasi-harmonic vibrations decreases with
$\phi$, some particles still find pathways to travel large distances
and allow the structural relaxation of the system.

Several attempts have been previously made to empirically fit the 
non-Gaussian shape of $G_s(x,t)$ with known functional forms.
Weeks {\it et al.}~\cite{weeks} have tried to fit their experimentally 
measured $G_s(x,t)$ with 
a stretched exponential function, in order to fit both the broad 
tails and the narrow center. Attempts have also been made 
to fit both components 
of the van Hove function as the sum of two different 
Gaussian functions~\cite{kegel,gao}. 
However, neither attempts seem to give 
satisfactory results since the shape of the distribution changes with time.
Basing their analysis on numerical
simulations of a Lennard-Jones system, Stariolo and Fabricius~\cite{stariolo}
recognized that the tails are probably 
better fitted with an exponential function in some time window.
Using extensive data, it has recently been shown~\cite{pinaki} that the 
$G_s(x,t)$, for different glass-formers, colloidal hard spheres
and granular materials close to jamming are better represented by a 
superposition of a central Gaussian along with an 
exponential tail for the large distances, which crosses over, 
at large times, to a Gaussian form. The model therefore allows one 
to describe the data at different times without changing 
the fitting formula in the middle of the game.
As we show below, the exponential tail is interpreted as the 
direct consequence of the occurrence of 
rare events of particles undergoing large displacements
that are statistically distributed. 
                                                           
\section{Random walk analysis}
\label{modelsec}

\subsection{Modeling single particle dynamics}

There have been several attempts to map the
heterogeneous single particle 
dynamics of gels and glass-formers
to some stochastic process. 
Closely related to our approach are the ones of 
Refs.~\cite{odagaki,monthus,epl}.
Odagaki and Hiwatari~\cite{odagaki} have studied the 
dynamics of atoms near the glass transition of simple classical 
liquids, on the basis of a  mesoscopic stochastic-trapping diffusion model, 
and calculated various dynamical quantities such as the 
mean squared displacement, non-Gaussian parameter and intermediate
scattering functions.
Monthus and Bouchaud~\cite{monthus} have looked at various models of 
independent particles hopping between energy traps which have
relaxation functions similar to glass-formers. They also show 
that diffusion in trap models can be described 
using the formalism of the continuous time random walk (CTRW)~\cite{mw}, 
widely used in many different areas of physics.
Finally, Berthier {\it et al.}~\cite{epl} also proposed to describe
the process of self-diffusion in glass-forming materials in terms
of a CTRW picture, and they base their analysis on the study 
of spin facilitated models. In this context, the CTRW picture
directly follows quite generically
from the spatially heterogeneous nature of the 
dynamics~\cite{jung1} .

\begin{figure}
\begin{center}
\psfig{file=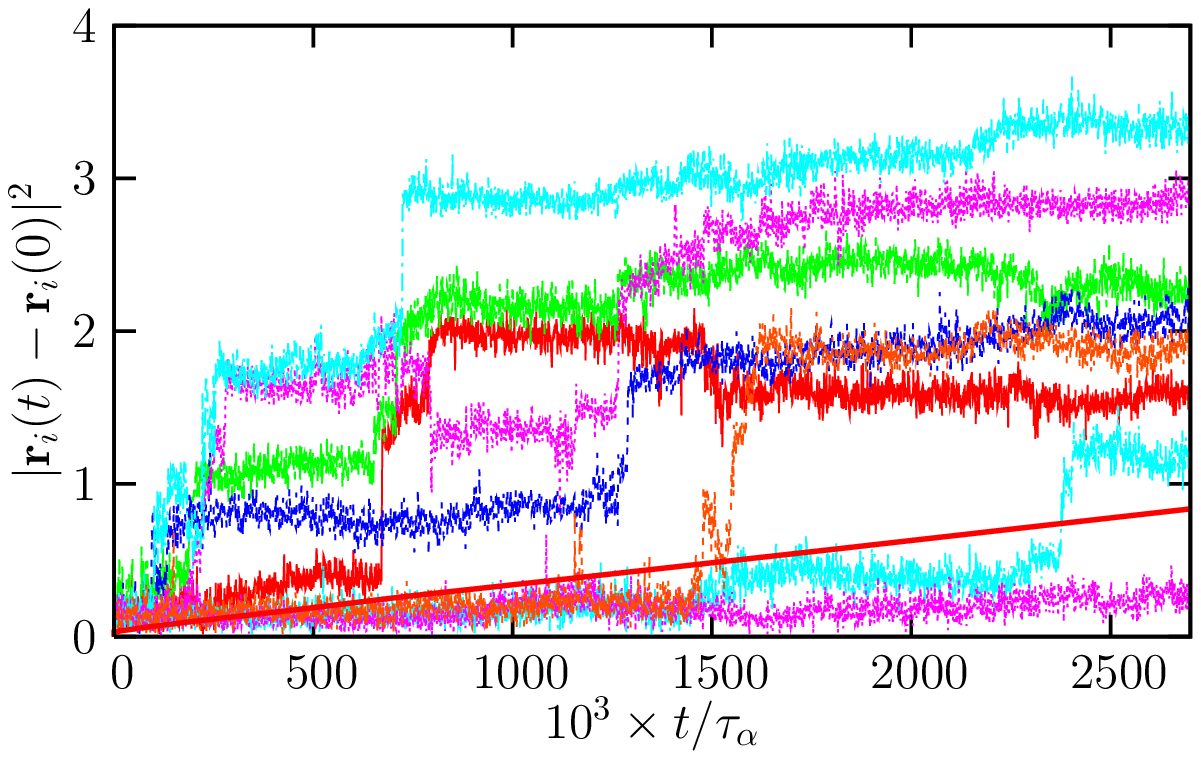,width=7cm}
\psfig{file=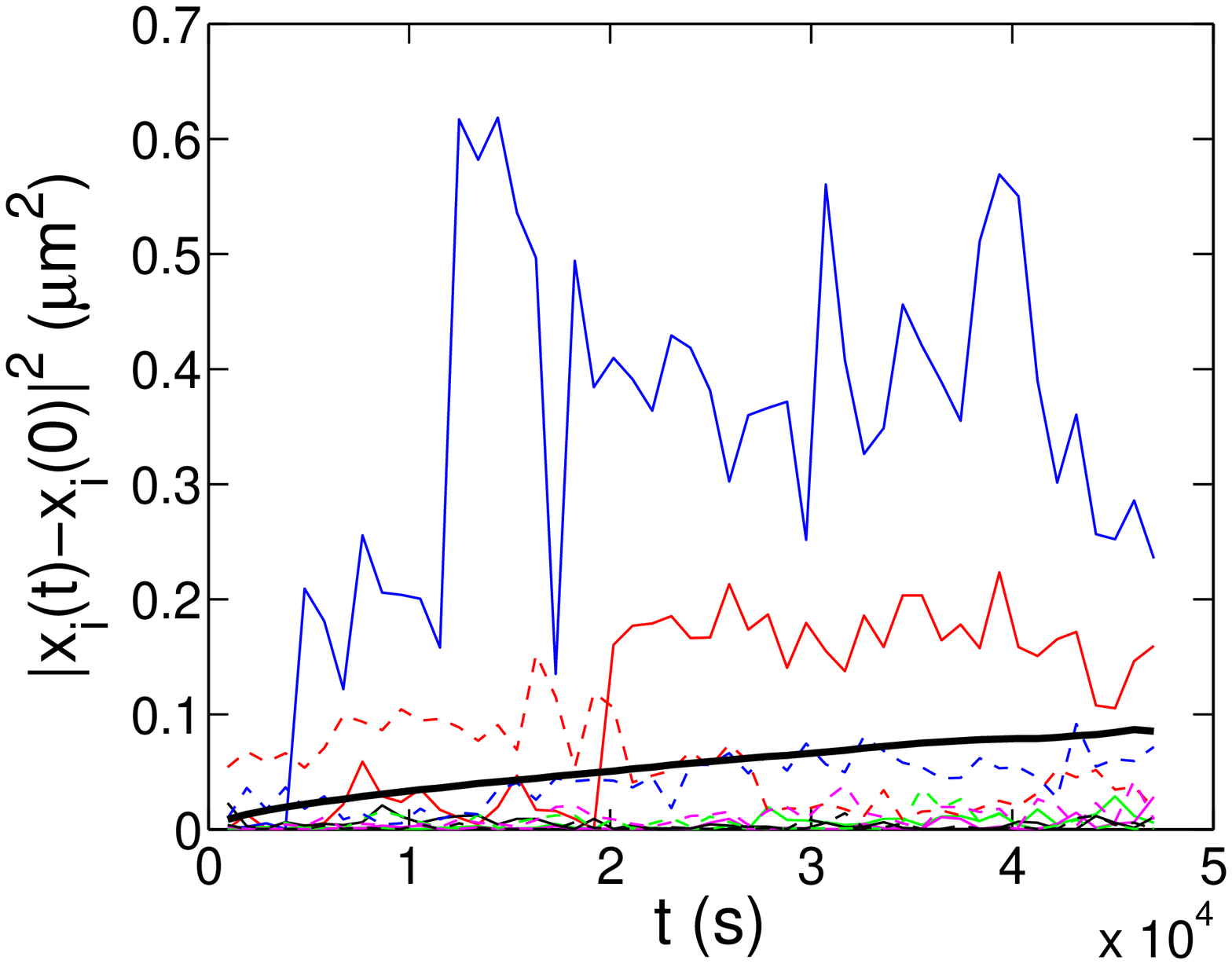,width=8.5cm}
\end{center}
\caption{Temporal evolution of squared displacement from an initial 
position for (top) particles in a binary Lennard Jones liquid 
at low temperature, $T=0.435$ (top), 
and (bottom) for the attractive colloidal particles at $\phi=0.429$.
While some particles rattle around their mean position, others perform one
or several quasi-instantaneous jumps. 
Occurrence of jumps occur randomly in time
and are random in size. 
The straight line in the plots 
corresponds to the mean-squared displacement.}
\label{ctrwfig}
\end{figure}

In fact, a convincing empirical 
rationale for this type of approaches stems from a 
visual inspection of particle trajectories in materials 
with slow dynamics, such as the ones 
shown in both panels of Fig.~\ref{ctrwfig}, 
which represent examples of particle displacements for a Lennard-Jones 
supercooled liquid~\cite{lj} and for the present colloidal system. 
Direct visualization reveals that, when observed on a timescale
comparable to $t^\star$,  
most of the particles simply perform a large number 
of localized vibrations around their initial position, 
just as in a disordered solid. However, 
the particles that contribute to the tail of the van Hove function 
undergo one or several quasi-instantaneous jumps separating long periods of 
localized vibrations, as can be seen in Fig.~\ref{ctrwfig}. 
For both systems, we observe that these jumps occur randomly in 
time and also have distributed amplitudes. Therefore,
a continuous time random walk~\cite{mw} should be a 
good coarse-grained stochastic model 
for single particle trajectories of these systems as suggested 
before~\cite{odagaki,monthus,epl,jung1}.

The case of very low density gels is peculiar, 
since in these systems there is a well-defined network of quasi-immobile 
particles existing along with the free particles. Therefore,  
the system can truly be decomposed into two dynamically distinct 
families, whose properties directly follow from the heterogeneous
nature of the stucture of these gels. 
Indeed, a two-family dynamical model has been shown to fit the 
van Hove distribution functions obtained in 
computer simulations of gel systems for a wide
window of parameters~\cite{pablo}. However, for denser systems, 
we have no structural basis to assume that such a distinction
can be made, although this has been done in other studies~\cite{langer}.
For supercooled liquids, it can even be quantitatively 
established by simulations~\cite{rob}
that dynamic heterogeneity at the particle scale 
has no such structural origin. 
Since the present system lies somewhat in between
low density gels and dense glasses, it is not obvious 
{\it a priori} whether 
we should adopt a glass (one family) 
or a gel (two families) description. 
In fact, we will show that the strong hypothesis of a two-family 
model is not necessary to account for our measurements. Therefore we will  
model the system as a collection of indistinguishable
particles undergoing continuous time random walks and we
prove below that such a modeling accounts well for the data presented 
in Fig.~\ref{fit}. Obviously, a two-family model would also 
fit our data very well since one can always artificially 
separate one group of particles into two distinct subgroups, 
(the reverse is not necessarily true). 

\subsection{A simplified CTRW model}

We now describe the CTRW model, introduced 
in Ref.~\cite{pinaki}, which will be used 
to fit the experimental data. We consider particles 
undergoing a stationary, three-dimensional, isotropic 
random walk process, as in the 
original Montroll-Weiss CTRW model~\cite{mw}, and add to the process 
localized vibrations occuring on a fast timescale in between the jumps.
We assume that vibrations are Gaussian and distributed 
according to 
$f_{\rm vib}(r) = 
(2 \pi \ell^2)^{-3/2} 
\exp(-r^2/2 \ell^2)$, so that $\ell^2$ represents 
the variance of  the size of vibrations.
%the root mean size of the vibrations.
We also assume that the jump size is distributed according to 
$f_{\rm jump}(r) = (2 \pi d^2)^{-3/2}
\exp(-r^2/2 d^2)$, introducing $d^2$, 
the variance in the size of the jumps.
%the root mean size of the jumps. 
The last ingredient needed to define the CTRW model is the distribution
of times between jumps, called the waiting time distribution~\cite{mw}, 
which we denote as $\phi_2(t)$, 
for reasons that will become clear in a moment. 

With these definitions, one can express the van Hove function as~\cite{mw} 
\begin{equation}
G_s(r,t) = \sum_{n=0}^{\infty} p(n,t) f(n,r),
\label{sum}
\end{equation} 
where $p(n,t)$ is the probability to make $n$ jumps 
in a time $t$, and $f(n,r)$ is the probability to move a distance 
$r$ in $n$ jumps~\cite{mw}. These probabilities involve convolutions 
and are more easily expressed in the Fourier-Laplace domain, 
($r,t$) $\to$ ($q,s$). The sum in Eq.~(\ref{sum}) is geometric and can
be performed easily to yield the well-known result~\cite{tunaley}:
\begin{equation}
G_s(q,s) = f_{\rm vib}(q) \frac{1-\phi_1(s)}{s}  + 
f(q) f_{\rm vib}(q) \frac{\phi_1(s)}{s} \frac{1-\phi_2(s)}{1-\phi_2(s)
f(q)} ,  
\label{model}
\end{equation}   
where we defined $f(q) \equiv f_{\rm vib}(q) f_{\rm jump}(q)$
and the distribution $\phi_1(t)$ is related 
to the waiting time distribution $\phi_2(t)$ through 
the Feller relation~\cite{tunaley}:
\begin{equation}
\phi_1 (t)=\frac{\int_t^\infty{dt'}\phi_2 (t')}{ 
\int_0^\infty dt' \phi_2(t') t'}. 
\label{feller}
\end{equation}
Physically, $\phi_1(t)$ represents the distribution
of the time, $t$,  a walker takes to undergo a jump starting 
from an arbitrary initial condition at time $t=0$. 
Note that $\phi_1$ becomes equal to $\phi_2$ when the distribution
of waiting time is a simple exponential, while it also follows
that the moments of $\phi_1 (t)$ are larger than those of 
$\phi_2 (t)$ if the distributions are broader than 
exponential~\cite{barkai1, barkai2, jung2}.
In a measurement of the van Hove function, $\phi_1(t)$ represents the 
distribution of the time to the first observed jump, 
as seen in the first term in the right hand side of Eq.~(\ref{model}), 
corresponding to $n=0$ in Eq.~(\ref{sum}). The distribution
$\phi_2(t)$ quantifies the time between subsequent jumps
and contributes to the second term in Eq.~(\ref{model}) which 
contains the contribution of all the terms with $n>0$ in the sum
(\ref{sum}).

The importance of the distinction 
between the first and subsequent jumps
in order to derive the 
correct expression of the van Hove function
was emphasized long ago by Tunaley~\cite{tunaley}, and is crucial 
when the distribution of waiting time becomes broad. 
As noted by Monthus and Bouchaud~\cite{monthus}, 
and by Barkai and coworkers~\cite{barkai1,barkai2}, this first term 
is in fact directly responsible of the aging dynamics observed 
in CTRW characterized by ``fat'' waiting time distributions. 
In Ref.~\cite{barkai2},
Barkai {\it et al.} even provide an example of a system for which 
the average time to the first jump is infinite, while 
the average time between jumps is finite: Eq.~(\ref{model})
then shows that in that case 
particles never leave their initial positions. 
Jung {\it et al.}~\cite{jung1,lutz} 
refer to the two distributions as ``persistence'' and ``exchange'' 
and relate them to the decoupling 
phenomena observed in supercooled liquids.

To proceed further and use Eq.~(\ref{model}) to fit 
experimental or numerical data, one needs input about the
waiting time distribution. It has been claimed by Odagaki and 
Hiwatari ~\cite{odagaki}  
%in the beginning of the 90's 
that waiting time 
distributions in a binary mixture of soft spheres 
becomes fat with power law tails 
at low temperature, as in the trap models studied
of Monthus and Bouchaud~\cite{monthus}. 
Garrahan and coworkers
performed extensive studies of waiting time distributions 
both in kinetically constrained glass models~\cite{jung1,jung2} 
and more recently
using molecular dynamics simulations~\cite{lutz}. Their results clearly 
confirm that waiting time distributions in glass-forming
systems are not trivial. In particular, they report measurements
of various moments of the distributions $\phi_1$ and $\phi_2$ 
and confirm that they evolve differently with temperature~\cite{jung2,lutz}, 
establishing the complex nature of the waiting time distributions for 
glass-formers.

Using these insights, we have suggested~\cite{pinaki} the following
simplification to make the use of Eq.~(\ref{model}) 
much more practical. In the absence of 
definite information on the detailed shape of $\phi_2(t)$, 
we characterize $\phi_1(t)$ and and $\phi_2(t)$ in Eq.~(\ref{model})
by their respective first moments, $t_1$ and $t_2$, and we generally 
expect that 
\begin{equation}
t_2 \leq t_1.
\label{ineq}
\end{equation}
We assume that 
the distributions $\phi_1 (t)$ and $\phi_2 (t)$ are 
exponential, $\phi_1 (t) = t_1^{-1} \exp(-t/t_1)$ and 
$\phi_2 (t) = t_2^{-1} \exp(-t/t_2)$, and that they are 
independent from one another. The real link between them 
in the Feller relation (\ref{feller}) and their complex shapes 
are now hidden in the inequality (\ref{ineq}).

\subsection{The exponential tail}

\begin{figure}
\begin{center}
\vspace*{0.45cm}
\psfig{file=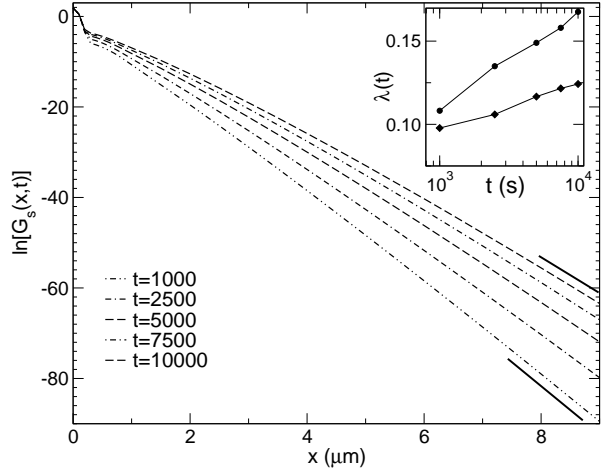,width=7.cm,angle=-90}
\end{center}
\caption{\label{fig3} Self-part of the van Hove function predicted by the 
model in Eq.(\ref{model}) with parameters 
$t_1=3 \times 10^5$~s, $t_2 = 10^4$~s, 
$\ell = 0.08 {\mu}m$ and $d = 0.284 {\mu}m$ at different times $t$. 
We show the data on an extended vertical scale 
to show that the tail is indeed very close to being 
exponential. Inset: Fitted slope $\lambda$ for different 
precision measurements, $G_s(r,t){\approx}10^{-5}$ (top) and 
$G_s(r,t){\approx}10^{-30}$ (bottom). The very slow growth simply reflects 
crossover towards the long-time Gaussian form of the distribution 
at fixed $x$.}
\end{figure}

The first term in Eq.~(\ref{model}), which  corresponds to the particles 
undergoing localized vibrations modulated by the waiting time 
distribution for first jumps, controls
the shape of the central part of the van Hove function $G_s(x,t)$. The 
second term in Eq.~(\ref{model}) is responsible for 
the broad tail in $G_s(x,t)$ and stems from particles 
which have performed one or several jumps 
after a time $t$. 
Using parameters relevant for our colloidal system (see below
for the details of the fitting procedure), we 
present on an extended vertical scale, the predictions
of Eq.~(\ref{model}) concerning the shape of the van Hove 
function and its evolution with time in Fig.~\ref{fig3}. 
The van Hove functions
can clearly be described as the superposition of ``mobile''
and ``immobile'' particles with broad tails that are well fitted by 
an exponential decay for large $x$:
\begin{equation}
G_s(x,t) \sim \exp \left(  -\frac{x}{\lambda(t)} \right), 
\end{equation}
which defines a new lengthscale $\lambda(t)$. 

In fact, a close to exponential decay of the van Hove function 
is present in the original CTRW model~\cite{mw} when 
distances outside the realm
of central limit theorem are considered. 
Using a saddle-point calculation, we have proved~\cite{pinaki} 
analytically that Eq.~(\ref{model}) generically leads
to broad distributions that indeed decay exponentially
(with logarithmic corrections).  
Interestingly this expansion can be obtained 
independently of the actual shape of the distributions,
establishing its universality. 
We have also shown~\cite{pinaki}
that these tails simply become enhanced in glassy materials, and 
are therefore more easily measured using typical experimental
accuracy. 

Using the exact solution from Eq.~(\ref{model}) shown in 
Fig.~\ref{fig3}, we fit 
the decay of $G_s(x,t)$ with an exponential 
function for two measurements of different precisions
corresponding to $G_s$ levels of $10^{-5}$ as in typical 
experiments, and of $10^{-30}$, which is obviously not accessible
experimentally. We find that the lengthscale 
$\lambda(t)$ slowly increases with time, 
the growth being slower for the most asymptotic measurements. 
This suggests that if one were to measure $\lambda(t)$ for even 
lower values of $G_s(x,t)$, $\lambda(t)$ would be almost constant,
in agreement with the saddle-point calculation. In fact, most 
of the time dependence of $\lambda(t)$ observed through fitting 
is due to the distribution
crossing over, at fixed $x$ and increasing $t$, to its long-time 
Gaussian limit. We conclude therefore that probably the ``growing lengthscale''
$\lambda(t)$ does not carry any deep physical information.  

Finally we remark that, quite often, 
the quantities ${4\pi}r^2 G_s(r,t)$ or even $P(\log_{10}r,t)
\propto r^3 G_s(r,t)$ are measured in simulations~\cite{puertas,kob},  
% plotted in numerical work
and the appearance of a secondary 
peak in $r$ at low temperature is given a large significance, supposedly
signalling the change towards an ``activated'' dynamics
with ``hopping'' processes. 
We would like to inform that within our CTRW model (which is a 
purely ``hopping'' model), a secondary peak is not necessarily present. 
Although the functions $r^2 \exp(-r/\lambda)$ and 
$r^3 \exp(-r/\lambda)$ describing the tails
have a maximum 
at some value of $r$, this peak is sometimes buried below the 
Gaussian central part 
of the van Hove function, so that only a shoulder (instead
of a secondary maximum) is observed.
A peak emerges, for instance, when the ratio between times 
$t_1$ and $t_2$ is large enough, the precise limiting value 
depending also on the parameters $d$ and $\ell$.  
Therefore, we believe that the observation of such peaks
is not in general indicative of a deep change in the physical
behaviour of the system.
 
\subsection{Fitting the data}

\begin{table}
\begin{center}
\begin{tabular}{|c|c|c|c|c|c|c|}
        \hline
        \hline
$\phi$ & $\ell$ & $d$ & $t_1^{\rm th}$ &     
$t_2^{\rm th}$ & $t_1^{\rm exp}$ & $t_2^{\rm exp}$ \\
        \hline 
0.37  & 0.14 & 0.195 & 300   & 70  & 59  & 36 \\
0.403 & 0.10 & 0.251 & 4000  & 800   & 955 &  448 \\
0.429 & 0.08 & 0.279 & 60000 & 5000   & 11050 & 4047 \\
0.440 & 0.06 & 0.284 & 300000 & 10000 & 22300 & 7086 \\
        \hline
        \hline
\end{tabular}
\caption{\label{table} Fitting parameters used to get the fits 
shown in Fig.~\ref{fit}. Timescales are in seconds, 
lengthscales in microns. }
\end{center}
\end{table}

We have used the model, given by Eq.~(\ref{model}), with the four fitting 
parameters $\{d, \ell, t_1, t_2 \}$ described above to fit 
the van Hove function $G_s(x,t)$ measured 
in our experimental system. Like in 
our previous work with different materials showing 
slow dynamics~\cite{pinaki}, suitable choice of the fitting 
parameters results in very good fits of the 
experimental data, as can be recognized from Fig.~\ref{fit}.
The fitting parameters we have used are presented in Table~\ref{table}.

To confirm that the good agreement obtained from the model
is not due to a large number of free parameters that would allow to fit 
any set of data, we have tried to compare our choice for the waiting times,
$t_1$ and $t_2$, with the same quantities being measured directly
from the observed trajectories.

To do so, we must determine the ``jumps'' from our trajectories.   
In our experiments, we say that a particle undergoes a jump 
if the magnitude of its displacement between two successive 
experimental frames is larger than a threshold, $x_{\rm cut}$. 
Here we consider displacements only in one dimension, and 
use $x_{cut} = 0.1~{\mu}$m, which is slightly larger than
the typical lengthscale for the vibrations, $x_{\rm cut} >
\ell$.
Similar to the CTRW model, we measure two distinct timescales 
associated with the jumps and separately record 
timescales to the first jump from an arbitrary initial condition, 
and timescales between jumps. Given that our experimental
trajectories have a finite duration, we observe particles 
which do not jump, meaning that we probably 
underestimate both timescales. 
Moreover, it needs to be noted that 
in the present experiment, only trajectories where at least 
two jumps have occured are being recorded meaning that
$t_1$ is slightly more underestimated than $t_2$ in our measurements.
From the statistics of the observed events, we obtain two
time distributions, from which we compute the first moments, 
which we label as $t_1^{\rm exp}$ and $t_2^{\rm exp}$, respectively.

\begin{figure}
\begin{center}
\vspace*{0.6cm}
\psfig{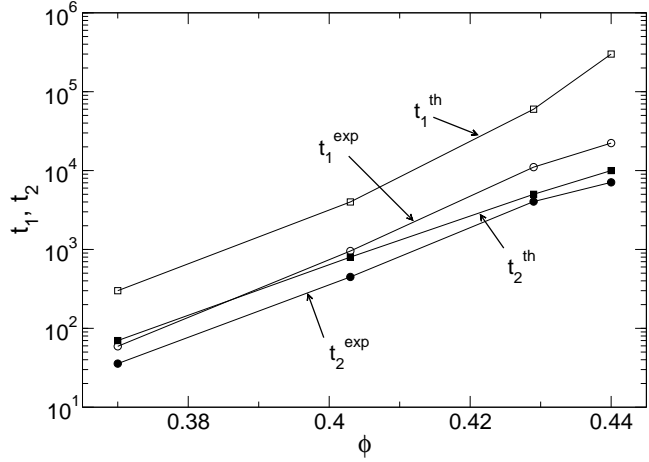}
\end{center}
\caption{\label{taudat}
The times $t_1$ and $t_2$ obtained directly from experiments ``exp''
are compared to the timescales obtained through the fitting procedure ``th''.
The $\phi$ dependence of both sets of data is similar, and the 
inequality $t_1 > t_2$ is strong in both cases, indicative of 
a broad distribution of waiting times $\phi_2(t)$.}
\end{figure}

We can then compare the experimental data
to the results obtained through the fitting procedure, 
which we label as $t_1^{\rm th}$ and $t_2^{\rm th}$, as shown 
in Fig.~\ref{taudat}. We find that the average waiting times, 
measured directly from 
experiments and by using the CTRW fitting procedure, show 
very similar trends when volume fraction is varied.  
This good agreement gives evidence that our modeling of 
the dynamics is physically correct, and that our fitting procedure
of the van Hove function indeed yields a detailed 
statistical information on the particle trajectories.

However, the numbers for $t_1$ and $t_2$ obtained from the 
CTRW model are higher than the numbers extracted from 
experimental measurements by a factor 2 and 10, respectively.
There can be several reasons for this mismatch, which might originate 
from the model or from the experimental determination
of waiting times, or from both.
The waiting time distributions are perhaps far more complex 
than the exponential distributions that are used in our model. 
But, as mentioned above, we have good reasons to believe 
that waiting times are slightly underestimated in our 
experimental analysis, 
$t_1$ more than $t_2$, a trend compatible with Fig.~\ref{taudat}.
One could also imagine the presence of back and forth motions,  
as seen in Fig.~\ref{ctrwfig}, and that would  erroneously be
counted as jumps, again biasing the experimental waiting times
towards small values, in agreement with the results 
presented in Fig.~\ref{taudat}.  
Given these possible sources of discrepancy,  
we conclude that the agreement reported 
in Fig.~\ref{taudat} is quite satisfactory.  

\section{Conclusion}
\label{conclusion}

In this paper, we have analyzed the heterogeneous dynamics
of a colloidal system which undergoes dynamical arrest 
at a volume fraction intermediate between low density 
gels and dense glasses. 
We have focused our attention on single particle 
trajectories and have analyzed in detail the 
self-part of the van Hove distribution functions. These 
distributions are strongly non-Gaussian with tails 
that are broad and decay close to exponentially 
with distance. We have shown that a simple 
continuous time random walk analysis proposed
in the context of glass and jamming transitions 
describes the experimental data in a very satisfactory
manner, showing that the present experimental system 
shares deep similarities with other glassy systems.

\ack We thank J.  R\"ottler and M. Kennett for inviting us to 
participate to the workshop ``Mechanical behaviour in glassy behaviour'', 
Vancouver, July 21-23, 2007, which led to the present collaborative work.
We would also like to thank David Reichman and Andreas Heuer 
for useful correspondence.
Financial support from the Joint Theory Institute
(Argonne National Laboratory and University of Chicago),
CEFIPRA Project 3004-1, and ANR Grants TSANET and DYNHET 
is acknowledged.

\end{document}